\documentstyle [12pt,epsf]{article}

\def\Vcl{V_{\rm cl}}
\def\c{_{\rm c}}
\def\tr{{\rm tr}}
\def\Tc{T\c}

\def\eps{\epsilon}

\def\mh{m_{\rm h}}
\def\mw{M_{\rm W}}

\def\LE{{\cal L}_{\rm E}}

\def\gsim{ \,\, \vcenter{\hbox{$\buildrel{\displaystyle >}\over\sim$}} \,\,}

\def\num{\#}
\def\alphaw{\alpha_{\rm w}}

\def\skipl#1{
   \dimen1=\baselineskip
   \multiply\dimen1 by #1
   \vspace{\dimen1}
}
\def\skiplStar#1{
   \dimen1=\baselineskip
   \multiply\dimen1 by #1
   \vspace*{\dimen1}
}

\advance\voffset by -1.6in
\advance\hoffset by -0.4in

\def\sec#1{
   \skipl{1}
   \noindent
   {\bf #1}
   \skipl{1}
}

\begin {document}

\begin {flushright}
UW-PT-94-04
\end {flushright}
\skiplStar{10}
\noindent
{\bf AN OVERVIEW OF THE $\epsilon$ EXPANSION AND THE ELECTROWEAK
     PHASE TRANSITION%
\footnote{
   Talk presented at the NATO Advanced Workshop on Electroweak Physics
   and the Early Universe: Sintra, Portugal, 1994.
}%
}

\skipl{3}
\noindent\hspace*{1in}Peter Arnold

\skipl{1}
\noindent\hspace*{1in}Department of Physics, FM-15

\noindent\hspace*{1in}University of Washington

\noindent\hspace*{1in}Seattle, WA 98195

\noindent\hspace*{1in}U.S.A.

\skipl{3}
In this talk I'm going to summarize work done with Larry Yaffe
\cite{Arnold&Yaffe} on applying $\eps$ expansion techniques to the
electroweak phase transition.  As I shall review, analysis based on
the perturbative expansion of the effective potential, which has
already been discussed in this conference, is only valid
(in the minimal standard model)
when the zero-temperature Higgs mass is small compared to the
zero-temperature W mass.  The goal of my talk will be to discuss
what can be done in the experimentally more realistic case when
this condition fails.  I shall discuss how one might compute various
parameters of the transitions such as the correlation length at the
critical temperature, the latent heat of the transition, the bubble
nucleation rate, and the baryon violation rate after the transition.
Note that these are all quasi-equilibrium quantities.  In particular,
I shall modestly not attempt to tackle the much harder problems
directly associated with the generation of baryon asymmetry on the
expanding bubble walls.

Throughout this talk I will work in a simple toy model: the minimal
standard model with a single Higgs doublet.  I do this because the
model has relatively few unknown parameters, and so it is relatively
easy to analyze and results are not obscured by complicated parameter
dependence.  I call it a toy model because I don't care to step into
the debate of whether the minimal model has enough
CP violation\cite{Farrar&Shaposhnikov, Gavela}.  For simplicity, I will also
be ignoring the Weinberg angle and focusing on a pure SU(2) Higgs
theory.

\sec{Why life is simple when m(Higgs)$_{T=0}$ $\ll$ M(W)$_{T=0}$}

Consider the classical Mexican-hat potential for the Higgs:
\begin {equation}
   V_0 \sim -\mu^2\phi^2 + \lambda\phi^4 \,.
\end {equation}
And now consider the free energy of the system at finite temperature
in the background of a Higgs field $\phi$.  One contribution will be from
the classical energy $V_0$ above for that $\phi$.  But there will also be
contributions from the free energy of the real particles present in the
plasma.  Focus on the contribution from W bosons.  This contribution
(at one-loop) is nothing more than a formula you can look up in your old
graduate statistical mechanics book; it is the free energy of an ideal
Bose gas:
\begin {equation}
   \Delta F \sim T \int d^3k \, \ln\left(1 - e^{-\beta E_k}\right) \,,
\end {equation}
where the energy of a relativistic particle is
\begin {equation}
   E_k = \sqrt{\vec k^2 + \mw^2} \sim \sqrt{\vec k^2 + g^2 \phi^2} \,.
\end {equation}
Note that the W gets its mass from the background Higgs field, so that
$\mw \sim g\phi$.  Thus $\Delta F$ above depends on the background value
of $\phi$.  At high temperatures, one can expand $\Delta F$ in powers
of $\mw/T$, and one gets a high-temperature expansion of the form
\begin {equation}
   \Delta F \sim \num T^4 + \num \mw^2 T^2 - \num \mw^3 T + \cdots \,,
  \label{Delta F high T}
\end {equation}
where $\num$ indicates numerical constants that I'm not going to
bother writing down.

\begin {figure}
\vbox
    {%
    \begin {center}
	\leavevmode
	
	\epsfbox [140 250 500 550] {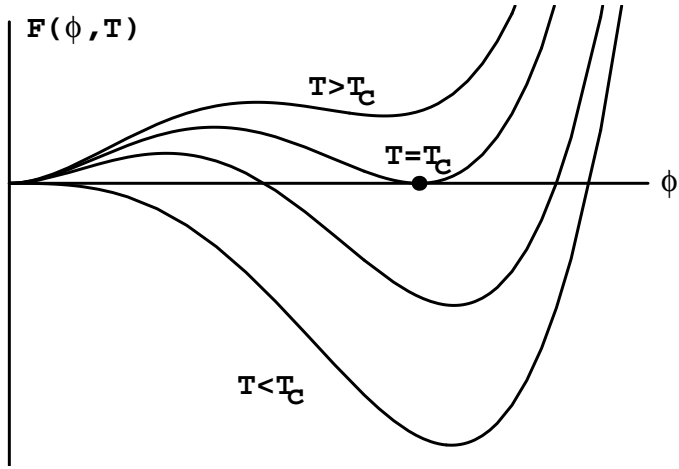}
    \end {center}
    \caption
	{%
	\label {figA}
        The form of the free energy, as a function of $\phi$, for
        different temperatures.
	}%
    }%
    \skipl{1}
\end {figure}

Putting the classical potential $V_0$ together with the W contribution
to $\Delta F$ gives%
\footnote{
  Of course, there are also scalar and other contributions to $\Delta F$.
  I'm just focusing on the W contributions here for the sake of brevity
  and pedagogy.
}
\begin {equation}
   F \sim V_0 + \Delta F
   \sim \hbox{const} + (-\mu^2 + g^2 T^2)\phi^2 - g^3\phi^3 T + \lambda\phi^4
                     + \cdots
   \label {free energy}
\end {equation}
(where I'm no longer going to bother even putting in the $\num$ signs).
The $g^2 T^2 \phi^2$ term comes from the $M^2 T^2$ term of the
high-temperature expansion of $\Delta F$.  This term is
responsible for the phase transition by turning the curvature of the
potential at the origin from negative at low temperature to positive at
high temperature.  The $g^3 \phi^3 T$ term, which comes from the
$M^3 T^2$ term, is responsible for making the phase transition first-order
as shown in fig.~\ref{figA}.  Specifically, there exists
a temperature $\Tc$ at which there are two degenerate minima of the
potential, and the expectation of $\phi$ changes discontinuously as one
cools through the transition.  I shall ignore the generally small effects
of subsequent terms in the high-temperature expansion (\ref{Delta F high T}).

Now consider the $T=\Tc$ curve in fig.~\ref{figA}.
The shape comes from the interplay of the $\phi^2$, $\phi^3$ and
$\phi^4$ terms in the free energy (\ref{free energy}): the $\phi^2$
term turns it up at the origin, the $\phi^3$ term turns it down again, and
the $\phi^4$ term turns it up at large $\phi$.  In particular, in the
region of $\phi$ near the symmetry-breaking minimum (where $\phi \not= 0$)
or near the hump in the potential, these three terms must be the same
order of magnitude, so that
\begin {equation}
   (-\mu^2+g^2T^2)\phi^2 \sim g^3 \phi^3 T \sim \lambda \phi^4 \,.
\end {equation}
{}From the last relation in particular, one finds that this region of $\phi$
is characterized by
\begin {equation}
   \phi \sim {g^3\over\lambda} T \,.
   \label {phi estimate}
\end {equation}

Now consider the order of the loop expansion parameter if we
perturbatively analyze the physics of the phase transition.  Each loop
costs a factor of $g^2$.  However, at high temperatures, each loop is also
associated with a factor of the temperature $T$, so that the cost of a loop
is actually $g^2 T$.  (One way to see this factor of $T$ is to remember
that, at finite temperature, Euclidean time becomes periodic with
period $\beta = 1/T$.  As a result, zero-temperature loop momentum
integrals $\int d^4 p$ are replaced by finite Fourier sums
$T \sum_{p_0} \int d^3 p$ over the frequencies, and a factor of $T$
then appears explicitly.)  Finally, by dimensional analysis,
$g^2 T$ must be divided by the mass scale $\mw$ of the problem to determine
the final, dimensionless cost of adding a loop.
The loop expansion parameter (after resummation) is then
\begin {equation}
   {g^2 T\over\mw} \sim {g T \over \phi} \sim {\lambda \over g^2} \,,
   \label{loop expansion}
\end {equation}
where I have used $\mw \sim g\phi$ and (\ref{phi estimate}).
$\lambda$ and $g^2$, however, are related to the zero-temperature Higgs
and W masses, so that the loop expansion parameter can be rewritten as
\begin {equation}
   {\lambda\over g^2} \sim
   { m^2(\hbox{Higgs})_{T=0} \over M^2({\rm W})_{T=0} }\,.
\end {equation}
So we see that the perturbative loop expansion is only valid when the
zero-temperature Higgs mass is small compared to the zero-temperature
W mass.  All of this power counting is reviewed in
ref.~\cite{Arnold&Espinosa}.

\sec{Why life is not simple}

\begin {figure}
\vbox
    {%
    \begin {center}
	\leavevmode
	
	\epsfbox [140 300 500 550] {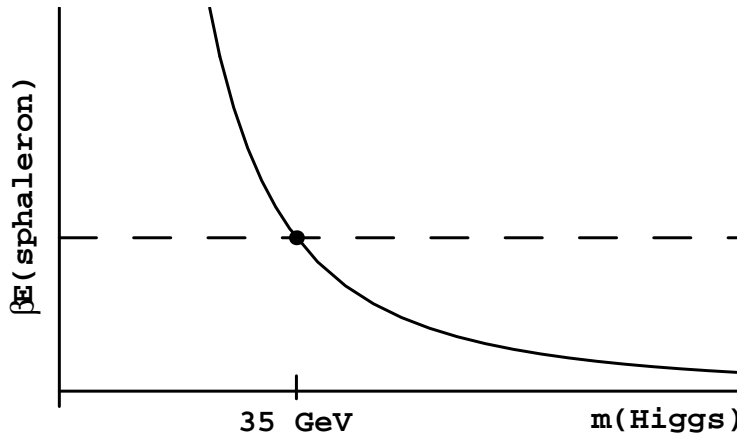}
    \end {center}
    \caption
	{%
	\label {figB}
        The Boltzmann exponent for baryon number violation vs.\ the
        zero-temperature Higgs mass.
	}%
    }%
    \skipl{1}
\end {figure}

One of the constraints of electroweak baryogenesis is that there must not be
any significant amount of baryon number violation after the completion of the
phase transition.  Otherwise, any baryon asymmetry generated during the
transition would be washed away to zero.  The rate of baryon number violation
is exponentially sensitive to the sphaleron mass:
\begin {equation}
   \hbox{rate} \sim e^{-\beta E(\hbox{sphaleron})}
   \sim e^{-\num \mw/g^2 T}
   \sim e^{-\num g^2/\lambda} \,.
\end {equation}
The second equality above comes because the sphaleron mass is of order
$\mw/\alphaw$, and the final equality because the exponent $\mw/g^2 T$
is (not coincidentally) nothing other than the inverse of the loop
expansion parameter (\ref{loop expansion}), which we previously determined
for the broken-symmetry phase.  You can now see that the
rate will be small after the transition only if the zero-temperature Higgs
mass (and hence $\lambda$) is small.  The inverse relationship between
the rate exponent and the zero-temperature Higgs mass is shown
schematically in fig.~\ref{figB}.  Requiring that the rate be small
compared to the expansion rate of the universe puts a lower bound on
$\beta E(\hbox{sphaleron})$ and hence an upper bound on the Higgs mass.
A careful analysis, based on the one-loop effective potential, yields
an upper bound on the Higgs mass of 35--40 GeV due to this
constraint\cite{Dine}.  This seems to rule out electroweak
baryogenesis in the minimal standard model.

\begin {figure}
\vbox
    {%
    \begin {center}
	\leavevmode
	
	\epsfbox [150 250 500 525] {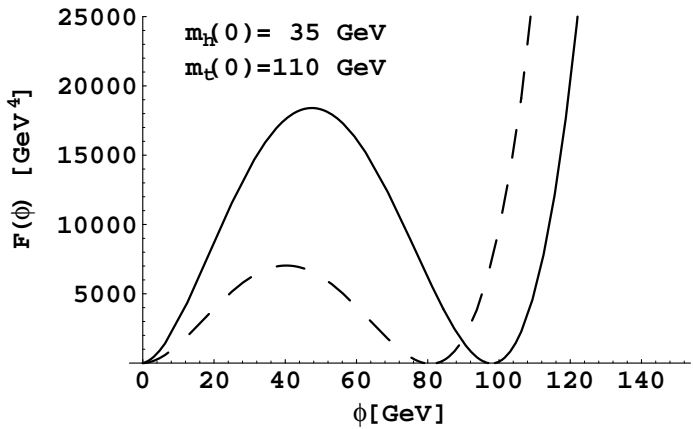}
    \end {center}
    \caption
	{%
	\label {figC}
        The (resummed) two-loop vs.\ one-loop free energy, as functions
        of $\phi$,
        for m(Higgs)$_{T=0}$ = 35 GeV at the critical temperature
        (as determined individually for each potential).
	}%
    }%
    \skipl{1}
\end {figure}

But now it's time to wonder whether a one-loop analysis can be trusted.
In particular, is a Higgs mass of 35 GeV in fact small compared to the
W mass of 80 GeV?  That depends on all the factors of 2 left out of the
order-of-magnitude analysis of loop corrections that I presented above.
To settle this issue, the two-loop potential was computed in
ref.~\cite{Arnold&Espinosa} (and the most dominant pieces independently in
ref.~\cite{Bagnasco&Dine}).  Fig.~\ref{figC} shows the difference between the
one-loop and two-loop potentials at the critical temperature.  Whereas
the VEV changed only by about 20\%,%
\footnote{
   Beyond leading order the VEV is an unphysical and gauge-dependent
   quantity.  The result shown here is for Landau gauge.
}
the height of the hump changed by
a factor of three!  The moral is that, for some quantities,
{(35 GeV)/(80 GeV)} is {\it not} a small number.  And so it is not
clear, in particular, whether higher-loop corrections to the baryon
number violation rate will be small.

But what can we do when mean field theory ({\it i.e.}\ perturbation
theory) fails---that is, when fluctuations around the mean field
$\phi$ are large and cannot be treated perturbatively.  To phrase
it another way, what do we do when the correlation length $1/\mw$
becomes very large at $\Tc$ so that the loop expansion parameter
(\ref{loop expansion}) is large?  Condensed matter physicists have
long dealt with this problem in the context of second-order phase
transitions, where the correlation length is infinite at $\Tc$.
Our strategy is to borrow one of their
techniques---the $\eps$ expansion---which
has been very successful for some systems.%
\footnote{
  The idea of applying the $\eps$ expansion to the electroweak phase
  transition has been suggested previously by Gleisser and Kolb
  \cite{Rocky},
  March-Russel \cite{March-Russel}, and probably by others that I'm
  not familiar with.
}

\sec{The $\eps$ expansion}

The $\eps$ expansion consists of the following steps.

{\it Step 1.}
Recall that equilibrium quantities can generally be expressed in terms
of the partition function $Z = \tr\, e^{-\beta H}$.
The exponential $e^{-\beta H}$ looks
just like an evolution operator $e^{i H t}$ when the time $t = i \beta$
is imaginary.  Just like the evolution operator, $Z$ has
a path-integral representation, and the only difference with
standard zero-temperature path-integrals is that time only extends for
Euclidean time $\beta$:
\begin {equation}
   Z = \int [{\cal D}\phi]
       \exp\left( - \int\nolimits_0^\beta d\tau L_{\rm E} \right) \,.
\end {equation}
Also, the trace is implemented by integrating over all configurations
where the initial state is the same as the final state, which translates
into the periodic boundary condition $\phi(0,\vec x) = \phi(\beta,\vec x)$ on
the path integral.  (Fermions have anti-periodic boundary conditions.)

{\it Step 2.}
Since the Euclidean time dimension only has extent $\beta$, it disappears
in the high temperature limit $\beta\to 0$.  That is, if one is studying
(equilibrium) physics at distance scales $l$ large compared to $1/T$, then
the Euclidean time dimension decouples and the four-dimensional theory
is reduced to an effective three-dimensional theory of the static modes
($p_0 = 0$).  The effects of fluctuations ($p_0 \not = 0$) in the
Euclidean time direction decouple, as powers of $\beta/l$, except for
renormalizations of relevant operators.  For example, the three dimensional
mass and couplings are related to the original four-dimensional ones by
\begin {eqnarray}
   m_3^2 &=& m_4^2 + g^2 T^2 + \hbox{higher-order} \,,
\\
   g_3^2 &=& g_4^2(T) + \hbox{higher-order} \,.
\label {coupling}
\end {eqnarray}
The $g^2 T^2$ term in the mass is just the term I discussed earlier that
turns the curvature of the potential at $\phi=0$ from concave down at
low temperature to concave up at high temperature, and so drives the
phase transition.  In four-dimensional language, the phase transition is
achieved by varying $T$.  In three-dimensional language, that corresponds
to varying the mass $m_3^2$ from negative values up through positive values.

{\it Step 3.}
As we shall see, the three-dimensional theory is difficult to directly
solve in a well-defined systematic fashion.  The trick of the $\eps$
expansion is to solve instead a related class of theories by generalizing
the 3 spatial dimensions to $4-\eps$ spatial dimensions, replacing the
action in the usual way by
\begin {equation}
   S(d{=}3) \to \int d^{4-\eps}x \left[
     (D\phi)^2 + m^2\phi^2 + \mu^\eps\lambda\phi^4 + \cdots \right] \,.
\end {equation}
(Note that I am defining my couplings, such as $\lambda$, to be dimensionless.)

{\it Step 4.}
For $\eps \ll 1$, the theory turns out to be solvable perturbatively in
$\eps$ by using renormalization-group (RG) improved perturbation theory.
I'll discuss this is more detail in a moment.

{\it Step 5.}
After computing to a given order in $\eps$, return to the world of three
spatial dimensions by taking $\eps \to 1$.

\sec{Review of a success story: pure scalar theory}

\begin {figure}
\vbox
    {%
    \begin {center}
	\leavevmode
	
	\epsfbox [140 250 500 550] {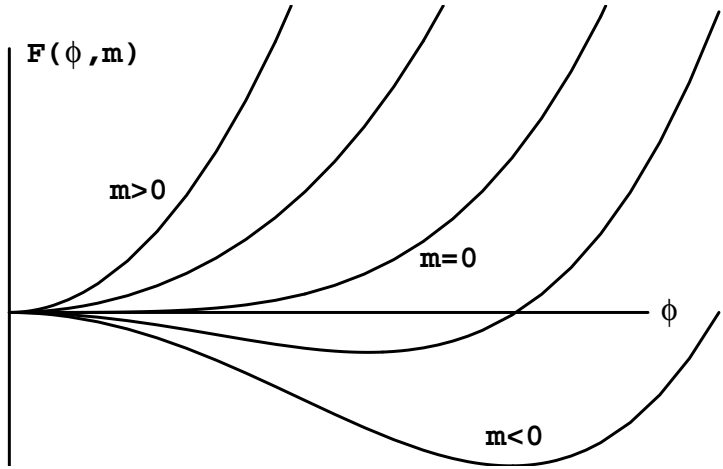}
    \end {center}
    \caption
	{%
	\label {figD}
        The classical potential for scalar theory, as a function of
        the mass of the effective 3 dimensional theory.
	}%
    }%
    \skipl{1}
\end {figure}

Forget about gauge theories for a moment and consider a theory of a
single, real scalar field:
\begin {equation}
   \LE \sim (\partial\phi)^2 + m^2\phi^2 + \mu^\eps \lambda\phi^4 \,.
\end {equation}
The phase transition of the tree-level potential is depicted
in Fig.~\ref{figD},
which shows a 2nd-order transition.  The 1-loop potential (appropriately
resummed) for this model turns out to predict a 1st-order phase transition,
but the loop expansion parameter turns out to be order 1 and so mean field
theory results cannot be trusted.%
\footnote{
  For a review, see section II.A of ref.~\cite{Arnold&Espinosa}.
}
In fact, this scalar theory goes by a familiar name near its critical point:
it is the Ising model, and the Ising model is known to have a 2nd-order phase
transition.  (More accurately, the scalar theory is in the same universality
class as the Ising model.  The field $\phi$ roughly corresponds to the Ising
spin averaged over a volume of size $1/\mu^3$.)

In a 2nd-order phase transition, the correlation length becomes infinite.
To study such a transition, one should therefore examine the system on
larger and larger distance scales and try to determine its scaling
behavior.  But, as particle physicists, we're all quite familiar with how
to change scale in $4-\eps$ dimensions.  We use the renormalization
group:
\begin {equation}
   \mu \partial_\mu \lambda = - \eps\lambda + c \lambda^2 + O(\lambda^2) \,.
\label {rng eq}
\end {equation}
If one set $\eps$ to zero, this would just be the usual four-dimensional
renormalization group for a pure scalar theory, with the one-loop
$\beta$-functions depicted explicitly as $c \lambda^2$.  $c$ is a
numerical constant that is easily computed.  Slightly away from
four dimensions, the additional term $-\eps\lambda$ on the right-hand side
simply represents the classical scaling of the interaction
$\lambda\phi^4$ due to its dimension.  For example, at tree level,
a change in $\mu$ by a factor of 2 can be compensated in
$\mu^\eps \lambda \phi^4$ by
changing $\lambda$ by a factor of $2^{-\eps}$.  This trivial
scaling is the sole content of the $-\eps\lambda$ term in (\ref{rng eq}).
It is the other terms which contain the physics of integrating out
degrees of freedom as the scale $\mu$ is changed.

\begin {figure}
\vbox
    {%
    \begin {center}
	\leavevmode
	
	\epsfbox [72 390 520 410] {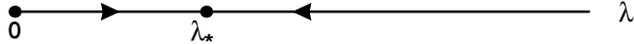}
    \end {center}
    \caption
	{%
	\label {figE}
        The renormalization group flow, into the infrared, of the pure
        scalar theory.
	}%
    }%
    \skipl{1}
\end {figure}

The flow of the RG equation (\ref{rng eq}) is shown in Fig.~\ref{figE}.
The arrows depict flow into the infrared ($\mu\to 0$).  At small
$\lambda$, the $-\eps\lambda$ term dominates and drives $\lambda$
bigger; at large $\lambda$, the $+c\lambda^2$ term dominates and
drives it smaller.  There is an infrared-stable fixed point when the
two terms balance at $\lambda_* = \eps/c$.  The long-distance behavior
of the system will be described by this fixed point.

When $\eps$ is small, the fixed point $\lambda_* = \eps/c$ will be
at small coupling, and therefore our neglect of two-loop and higher-order
contributions to the renormalization group in (\ref{rng eq}) will be
justified.  More generally, the loop expansion in $\lambda$ at the
fixed point is equivalent to an expansion in $\eps$.  We know, however,
that the loop expansion does not converge but is asymptotic, blowing up
at order $\lambda_*^n$ where $n \sim 1/\lambda_*$.  That means that the
expansion in $\eps$ will also be asymptotic, blowing up at order
$\eps^n$ where $n \sim 1/\eps$.  If we're lucky, $n \sim 1/\eps$ will
mean 3 or 4 when $\eps \to 1$ and the first few terms of the expansion
will be useful in three dimensions, giving a reasonable approximation
for quantities of interest.  If we're unlucky, $n \sim 1/\eps$ will
mean zero when $\eps \to 1$, and the asymptotic series will start
blowing up immediately after the first term and be useless in three
dimensions.

Are we lucky or unlucky?  Let's look at some examples.  The first is
the susceptibility exponent $\gamma$ defined by $\chi \sim |T-\Tc|^\gamma$
where $\chi$ measures the response of $\langle\phi\rangle$ to an
external source $J\phi$.  The result is
\begin {equation}
   \gamma = 1 + 0.167\eps + 0.077\eps^2 - 0.049\eps^3 + \cdots \,,
\end {equation}
which, for $\eps = 1$, gives $1.195+\cdots$.  This compares favorably to
the result $1.2405 \pm 0.0015$ which comes from more sophisticated
techniques (the $\eps$ expansion combined with Pade resummation, or
renormalization group studies directly in three dimensions).  It is
also consistent with lattice results, which have larger error bars.
It is also consistent with experiments, which have yet larger error bars.
I should also give a much more marginal example, which is the
correlation exponent $\eta$ defined by
$\langle \phi(r) \phi(0)\rangle \sim r^{2-d-\eta}$ as $r \to \infty$.
$\eta$ is simply twice the anomalous dimension of $\phi$ and is given
by
\begin {equation}
   \eta = 0.0185 \eps^2 + 0.0187 \eps^3 - 0.0083 \eps^4 + 0.0359 \eps^5
          + \cdots \,.
\end {equation}
If one adds the first three terms for $\eps=1$,
and ignores the fourth where the series
is clearly blowing up, one gets $0.0289.$
The actual result is believed to be about 0.035.
So the naive application of the $\eps$ expansion gives
an approximation to within about 20\%.

\sec{Electroweak theory}

The one-loop renormalization group equations for the couplings of
electroweak theory are similar to the case of scalar theory:
\begin {eqnarray}
   \mu\partial_\mu g^2 &=& - \eps g^2 + \beta_0 g^4 \,,
\\
   \mu\partial_\mu \lambda\phantom{{}^{2}} &=& - \eps g^2
                 + (a g^4 + b g^2 \lambda + c \lambda^2) \,.
\label {RG eqs}
\end {eqnarray}
The scalar $\beta$-function is a little more complicated now because
there are terms involving $g^2$ as well as $\lambda$.

\begin {figure}
\vbox
    {%
    \begin {center}
	\leavevmode
	
	\epsfbox [72 260 520 550] {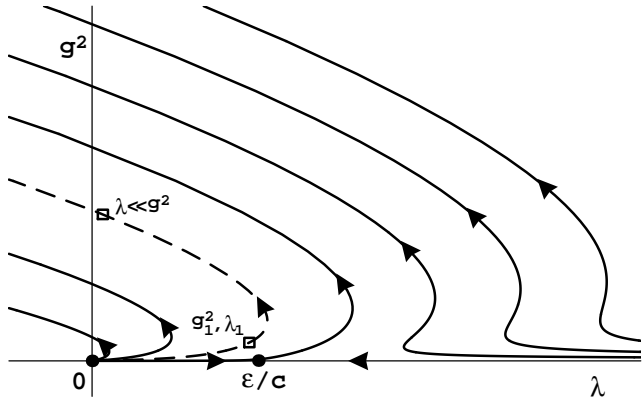}
    \end {center}
    \caption
	{%
	\label {figF}
        The renormalization group flow, into the infrared, of
        electroweak theory.
	}%
    }%
    \skipl{1}
\end {figure}

When one computes the numerical coefficients in these equations and
plots the renormalization group flow, one gets Fig.~\ref{figF}.
The $g^2=0$ line is similar to our previous result for pure
scalar theory.  Howvere, because of the flow in $g^2$, the fixed
point is no longer infrared stable.
As observed by Ginsparg \cite{Ginsparg},
there are in fact no infrared stable fixed points seen in the $\eps$
expansion.  This suggests that the phase transition might always be
1st-order.  Another, extremely hand-waving, suggestion of this comes
from the observation that all theories flow into the region $\lambda < 0$
and, at tree level, $\lambda < 0$ would be an unstable theory.  This
suggests that there is something suspect about flowing to arbitrarily
large distances, as one would desire if the transition were 2nd-order.

The fact that the transition is 1st-order can be made rigorous, for small
$\eps$, by remembering that we know how to solve the theory when
$\lambda \ll g^2$.  When $\lambda \ll g^2$, straightforward perturbation
theory is under control and predicts a 1st-order transition.  So now
consider a theory initially defined by the couplings $g_1^2$ and
$\lambda_1$ as depicted in Fig.~\ref{figF}.  The renormalization group
allows us to flow to larger distance scales and fins a completely
equivalent theory where $\lambda \ll g^2$, also depicted in the
figure.  This equivalent theory can then be solved with straightforward
perturbation theory and predicts a 1st-order transition.  In summary,
our method is as follows.

{\it Step 1.}
Start with initial couplings $g_1^2, \lambda_1$ at scale $\mu_1 \sim T$
(which is the scale where the Euclidean time fluctuations first decouple).

{\it Step 2.}
Use the renormalization group to flow to an equivalent theory with
$\lambda \ll g^2$.

{\it Step 3.}
We can then perturbatively compute the effective potential or whatever
other quantities are of interest.

{\it Step 4.}
Expand all result in $\eps$.

This procedure is closely related to the work of Rudnick \cite{Rudnick}
and Chen, Lubensky and Nelson \cite{Chen&Lubensky&Nelson}.%
\footnote{
   There is a long story about why the $\eps$-expansion
   prediction that the phase transition is always 1st-order has,
   historically, been believed to be incorrect in the U(1) case,
   where the theory models the phase transition of superconductors.
   For a discussion of problems with and loopholes to this belief,
   see section I.E of ref.~\cite{Arnold&Yaffe}.
}

In the first step, we started with initial couplings $g_1^2$ and
$\lambda_1$.  We know what the initial couplings should be in the
three dimensional theory; they are perturbatively related to the couplings in
the original 3+1 dimensional theory as in (\ref{coupling}).  But in
the $\eps$ expansion we have generalized the 3-dimensional theory to a
$4{-}\eps$ dimensional theory.  What couplings $g_1^2$ and $\lambda_1$
in the $4{-}\eps$ dimensional theory most naturally extrapolate to those
of the 3-dimensional theory as $\eps\to 1$?

This question is easily answered if one observes that the one-loop
renormalization group equations are independent of $\eps$ if
appropriately rescaled.  Specifically, take
\begin {equation}
  g^2 \to \eps \bar g^2 \,,
  \qquad
  \lambda \to \eps \bar\lambda \,,
  \qquad
  \mu \to \bar\mu^{-1/\eps} \,.
\label {rescale}
\end {equation}
The one-loop RG equations (\ref{RG eqs}) then become
\begin {eqnarray}
   \mu\partial_\mu \bar g^2 &=& - \bar g^2 + \beta_0 \bar g^4 \,,
\\
   \mu\partial_\mu \bar\lambda\phantom{{}^2} &=& - \bar g^2
                 + (a \bar g^4 + b \bar g^2 \bar\lambda + c \bar\lambda^2)
\end {eqnarray}
and are independent of $\eps$.  Thus, if I had simply plotted $g^2/\eps$
vs.\ $\lambda/\eps$ in Fig.~\ref{figF} rather than $g^2$ vs.\ $\lambda$,
the trajectories would have been ${\it independent}$ of $\eps$.
The natural generalization of any particular theory from 3 to $4{-}\eps$
dimensions is then clear: just keep the same place on the same trajectory.
So the relationship between the couplings is
\begin {equation}
       (g_{4{-}\eps}^2, \lambda_{4{-}\eps})
       = (\eps g_3^2, \eps \lambda_3) \,.
\end {equation}
Note that, even though the trajectories are independent of $\eps$
once the couplings are scaled by $\eps$, the rate at which those
trajectories are traversed is exponentially sensitive to $\eps$ as
a result of the scaling of $\mu$ in (\ref{rescale}).  In particular,
the trajectories are traversed {\it exponentially} slowly as $\eps\to 0$.

\sec{Example}

\begin {figure}
\vbox
    {%
    \begin {center}
	\leavevmode
	
	\epsfbox [140 250 500 550] {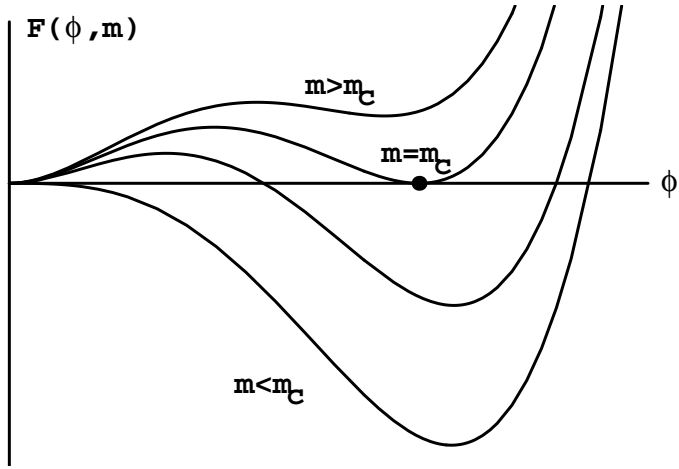}
    \end {center}
    \caption
	{%
	\label {figG}
        The Coleman-Weinberg potential as a function of $m$.
	}%
    }%
    \skipl{1}
\end {figure}

I'll now very roughly and schematically outline a sample calculation.
Consider the classical Higgs potential
\begin {equation}
     \Vcl \sim m^2 \phi^2 + \mu^\eps \lambda\phi^4 \,.
\end {equation}
Suppose I've used the renormalization group to run to
$\lambda \ll g^2$.  In particular, it's computationally convenient to
run to $\lambda = 0$.  I can now compute the 1-loop potential, and
the result for $\eps \to 0$ is the familiar Coleman-Weinberg potential:
\begin {equation}
     V^{{\rm (1-loop)}} \sim m^2 \phi^2
          + g^4 \phi^4 \ln\left(g\phi\over\mu\right) \,.
\label{V 1-loop}
\end {equation}
The dependence of this potential on the effective mass $m$ gives a 1st-order
phase transition, as shown in Fig.~\ref{figG}.  To study the system at the
critical temperature, I should find the critical value of $m$ where the two
minima are degenerate.  I can then compute, for example, the scalar
correlation length $\xi$ in the asymmetric phase at the critical temperature,
which at this order is related to the curvature of the potential.  One finds
\begin {equation}
    \xi^2 \sim {1\over V''(\phi_{\rm min})} \sim {1\over g^2\mu^2} \,.
\label {xi}
\end {equation}
[The dependence $1/g^2\mu^2$ is easy to obtain from the form of
(\ref{V 1-loop}) by simple scaling arguments.]
Now recall from the previous section that $g^2$ is $O(\eps)$ and that
the amount of running required to run to $\lambda = 0$ is exponentially
long as $\eps \to 0$, so that
\begin {equation}
   \mu \sim e^{O(1/\eps)} T \,.
\end {equation}
As a result, the dependence of (\ref{xi}) on $\eps$ is
\begin {equation}
   \xi^2 \sim {1\over\eps} e^{\#/\eps} {1\over T^2} \,,
\end {equation}
where $\#$ is computed from the renormalization group equations.
If one works to higher orders in $\eps$, one finds that the prefactor
of the exponential is a series in $\eps$:
\begin {equation}
   \xi^2 T^2 \sim {1\over\eps} e^{\#/\eps}
       (1 + \#\eps + \#\eps^2 + \cdots) \,.
\end {equation}
When I later refer to orders of the $\eps$ expansion, I shall be
referring to the expansion of the prefactor.  So ``leading-order''
will refer to a calculation that includes the first term of the prefactor,
``next-to-leading order'' includes the second term, and so forth.

\sec{Tests}

Now that I've described how to go about computing things with the $\eps$
expansion, it would be nice to find some cases where we can check to see
if the $\eps$ expansion is working.  One case where we {\it already}
know how to compute results is in the perturbative regime where the
initial $\lambda_1/g_1^2$ is $\ll 1$.  As a first check, we can
compute the results of the $\eps$ expansion in the same limit.
Table~\ref{tableb} shows the ratio of $\eps$ expansion results to the correct
answer in the limit $\lambda_1/g_1^2 \to 0$.  Note that leading-order in
$\eps$ is generally within roughly 50\% and next-to-leading order within
roughly 5-10\%.  (The exception is the last quantity, which is the difference
of free energies of the two ground states at the temperature where the
$\phi=0$ ground state is just loosing its meta-stability.)

\begin{table}
\begin {center}
\tabcolsep=8pt
\begin {tabular}{|l|cc|}             \hline
\multicolumn{1}{|c}{observable ratio}
  & LO          & NLO         \\ \hline
asymmetric phase correlation length
  & 1.14        & 0.94        \\
symmetric phase correlation length
  & 1.62        & 0.92        \\
latent heat
  & 0.77        & 1.04        \\
surface tension of a domain wall
  & 0.60        & 0.98        \\
$\Delta F(T_0)$
  & 0.24        & 0.56        \\ \hline
\end {tabular}
\end {center}
\caption
    {%
    \label {tableb}
    The ratio of the $\eps$-expansion results,
    computing prefactors through leading order (LO) and
    next-to-leading order (NLO) in $\eps$,
    to the corresponding three-dimensional result when $\lambda_1 \ll q_1^2$.
    }%
\end{table}

We have also tested the $\eps$ expansion against large $N$ results where
$N$ is the number of scalar fields in the theory.  I didn't have time to
discuss this in Sintra.  The result is that the $\eps$ expansion works
qualitatively but fails quantitatively at leading-order in $\eps$.
However, the $\eps$-expansion self-diagnoses its own quantitative
failure because a computation of next-to-leading order corrections
in $\eps$ shows that they are large compared to the leading-order results.

\begin {figure}
\vbox
    {%
    \begin {center}
	\leavevmode
	
	\epsfbox [150 250 500 500] {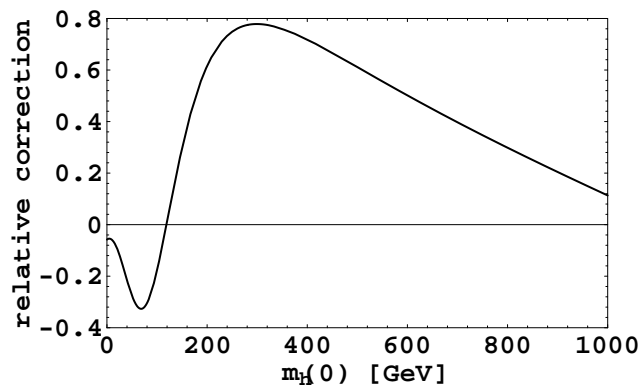}
    \end {center}
    \caption
	{%
	\label {figH}
        The relative size of the next-to-leading order correction compared to
        the leading-order result, as a function of zero-temperature
        Higgs mass.
	}%
    }%
    \skipl{1}
\end {figure}

The perturbative test discussed above is only of limited interest because,
after all, we already knew how to solve the theory in the perturbative
regime.  The large $N$ tests are of limited interest because we want
to know what happens when $N$ is small.  (Indeed, the large $N$ theory
is very different.  For example, it is not asymptotically free.)  What
we really want is a test of the $\eps$ expansion when
$\lambda_1 \gsim g_1^2$.  A natural thing to do is to check whether
the $\eps$ expansion diagnoses its own failure---that is, whether
next-to-leading order corrections are small or large compared to
leading-order results.  Computing results at next-to-leading order
in this regime is involved, and we only had the stamina to make one
test.  Our test quantity is related to the specific heat.  More
specifically, for technical reasons, it is
\begin {equation}
  (\hbox{asymmetric phase correlation length})^2
  \times (\hbox{latent heat}) \,.
\nonumber
\end {equation}
The calculation to next-to-leading order requires the 2-loop renormalization
group and a 2-loop computation of the potential.%
\footnote{
  If we had just computed the latent heat by itself, we would have
  needed the three-loop renormalization group, which is not currently
  known.  See ref.~\cite{Arnold&Yaffe} for details.
}
Fig.~\ref{figH} shows the results.

{}From this figure, we conclude that the $\eps$ expansion may be
quantitatively useful for $\mh$ below 150 GeV or so, where the correction
is within $\pm$30\%.  The expansion may be useful qualitatively, but
probably not quantitatively, for larger Higgs mass.  You should be cautioned,
however, that we have only tested one quantity in this way, and perhaps it's
better behaved than most.

\sec{Qualitative results for B violation}

The original motivation for this study was to examine the bounds on
electroweak baryogenesis, and so I should say a few words about the
implications of the $\eps$ expansion for the rate of B violation at
the completion of the phase transition.  In particular, one can
compare the results of renormalization-group improved perturbation
theory in $4{-}\eps$ dimensions to the results one would get with
naive, unimproved perturbation theory in $4{-}\eps$ dimensions.
At one loop, we find that RG improvement predicts a {\it stronger}
phase transition than the unimproved one-loop result, consistent with
the nature of the 2-loop correction of Fig.~\ref{figC} discussed
earlier.  One might assume that a stronger phase transition implies
a smaller rate of B violation after the transition.  Instead, we
find a {\it larger} rate of B violation in $4{-}\eps$ dimensions.
This conclusion is, as always, only rigorous for $\eps \ll 1$ and
may or may not hold qualitatively when $\eps = 1$.  (Ideally, one should
compute next-to-leading order corrections and see if they are under control.)
However, our result offers an important warning.  If one studies
the transition on a lattice, and finds a transition much stronger than
expected by perturbation theory, that does not necessarily mean
that baryogenesis is more viable in the model being studied.

I should conclude with a final caveat, however.  The calculation of the
B violation rate in the $\eps$ expansion is beset by additional technical
perils arising from the intrinsically 3-dimensional nature of the
sphaleron.  And so the extrapolation of $\eps\to 1$ may be even
more suspect than usual.  See ref.~\cite{Arnold&Yaffe} for details.

\end {document}